\documentclass[final,5p,times,twocolumn]{elsarticle}

\usepackage{epstopdf}
\usepackage{graphicx}
\usepackage{subfigure}
\usepackage{xcolor}
\usepackage{amsmath}
\usepackage{amssymb}
\usepackage{bbm}
\usepackage[thinc]{esdiff}

\usepackage{etex} 
\usepackage{array,arydshln} 
\usepackage{threeparttable} 
\usepackage{booktabs} 
\usepackage{colortbl}

\newcounter{LT@tables}
\newcounter{LT@chunks}

\newcolumntype{$}{>{\global\let\currentrowstyle\relax}}
\newcolumntype{^}{>{\currentrowstyle}}
\newcommand{\rowstyle}[1]{\gdef\currentrowstyle{#1}%
  #1\ignorespaces
}
\newcolumntype{L}[1]{>{\raggedright\arraybackslash}p{#1}} 
\newcolumntype{C}[1]{>{\centering\arraybackslash}m{#1}} 
\newcolumntype{R}[1]{>{\raggedleft\arraybackslash}p{#1}} 

\DeclareGraphicsExtensions{.pdf,.tikz,.eps}

\definecolor{blueThesis}{rgb}{0,.4,.8} 
\definecolor{darkblueThesis}{rgb}{0.0784314, 0.329412, 0.670588} 
\definecolor{lightblueThesis}{rgb}{0.623529, 0.807843, 1.} 
\definecolor{orangeThesis}{rgb}{.8,.4,.2} 
\definecolor{darkorangeThesis}{rgb}{0.6, 0.3, 0.15} 
\definecolor{lightorangeThesis}{rgb}{1., 0.807843, 0.623529} 
\definecolor{semilightorangeThesis}{rgb}{0.917647, 0.6, 0.462745} 

\renewcommand{\eqref}{\text{Eq. }\ref}
\newcommand{\fig}[1]{Fig. \ref{#1}}
\newcommand{\tab}[1]{Tab. \ref{#1}}

\newcommand{\e}{\mathrm{e}} 
\newcommand\gas{\text{g}}

\newcommand\bb{\text{bb}}
\newcommand\+{\oplus}
\renewcommand\-{\ominus}
\newcommand\rad{\text{rad}}
\newcommand\dd{\,\mathrm{d}}
\newcommand\kB{k_\mathrm{B}} 
\def\u#1{\,\mathrm{#1}} 
\newcommand{\Kn}{K\!n} 

\usepackage{natbib}
\bibpunct{(}{)}{;}{a}{}{,}

\journal{Journal of Aerosol Science 97 (2016) 22--–33}

\begin{document}

\begin{frontmatter}
	
	
	
	\title{Photophoresis on particles hotter/colder than the ambient gas in the free molecular flow
	}
	
	
	\author{C. Loesche, G. Wurm, T. Jankowski, M. Kuepper}

	\address{Fakult{\"a}t f{\"u}r Physik, Universit{\"a}t Duisburg-Essen, Lotharstr. 1, 47048 Duisburg, Germany}
	
\begin{abstract}
Aerosol particles experience significant photophoretic forces at low pressure.
Previous work assumed the average particle temperature to be very close to the gas temperature.
This might not always be the case.
If the particle temperature or the thermal radiation field differs significantly from the gas temperature (optically thin gases), given approximations overestimate the photophoretic force by an order of magnitude on average with maximum errors up to more than three magnitudes.
We therefore developed a new general approximation which on average only differs by 1 \% from the true value.
\end{abstract}


\end{frontmatter}


\section{INTRODUCTION}

If particles are entrained in a gaseous environment they are subject to photophoretic forces \citep{Yalamov1976_photohoresis_fm,Yalamov1976_photohoresis_co}.
Photophoresis is strongest for particles in a size range comparable to the mean free path of the surrounding gas. 
It is considered to act on dust in Earth's atmosphere \citep{Hidy1967,Yalamov1976_photohoresis_fm,Yalamov1976_photohoresis_co,Beresnev2003Levitation}.
Also, it might work in protoplanetary disks on particles as large as meter \citep{Krauss2005, Kuepper2014a}.
It can also aid to levitate particles in laboratory settings \citep{vanEymeren2012}.

In all applications the force can be estimated by analytical approximations which exist in the literature for the free molecule regime (\textit{fm}) \citep{Hidy1967,Yalamov1976_photohoresis_fm,Beresnev1993}, the continuum regime \citep{Yalamov1976_photohoresis_co} and the transition regime \citep{Reed1977, Mackowski1989aerosol}.
In the \textit{fm} regime, the interaction of gas molecules with a particle can be treated as individual collisions and we restrict our work to this case here. 

Previous approximations for the free molecule regime assume spherical particles that are suspended in a gas with its temperature only slightly deviating from the particle's surface temperature.
This is not always an appropriate assumption and the motivation of this work is to introduce an equation with an extended scope to describe photophoresis also for particle surface temperatures largely differing from the gas temperature, for instance by a factor of two. This is e.g. the case for laser induced photophoresis in the laboratory \citep{Daun2008c, Loesche2014, Kuepper2014b, Wurm2010}. In some of the experiments particles are embedded in a gas at room temperature and are illuminated with a laser of several $\u{kW\,m^{-2}}$, which heats the particles to several hundred K above room temperature. Also cool dust might be embedded in a hot gas environment in protoplanetary disks with temperature differences of an order of magnitude \citep{Akimkin2013}.
Additionally, in the optically thin parts of protoplanetary disks the gas temperature is different from the thermal radiation (0 K).
Comparison of the given approximations with numerical calculations for the free molecule regime showed that the classic approximations might deviate from the true value by more than an order of magnitude \citep{Loesche2012,ChrisDiss,Loesche2015LPI}.
We therefore suggest a more accurate analytic equation which also includes a significantly extended scope of gas temperatures and incorporates thermal radiation.
\section{PHOTOPHORESIS IN THE FREE MOLECULAR FLOW}\label{sec:fm_phothophoresis}
\begin{figure}[ht!]
	\centering
	\includegraphics{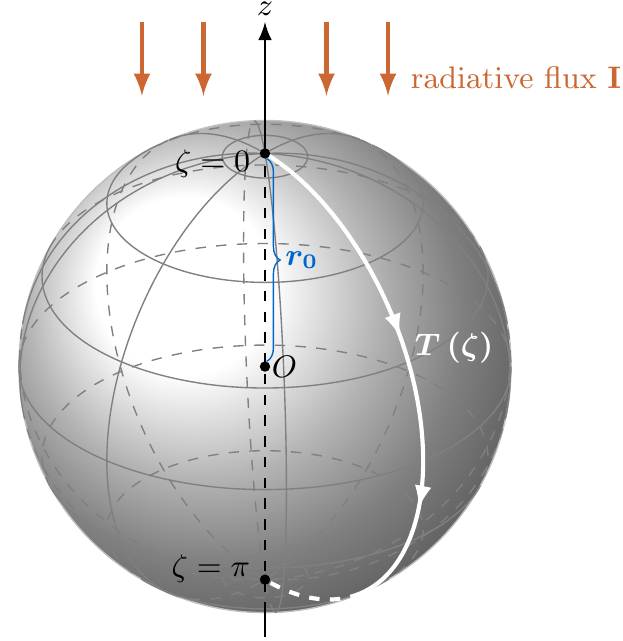}
	\caption{\label{fig:sphere_with_Trot}
		Visualization of the situation considered.
		Illumination is directed along $z$-axis, thus for a homogeneous particle the surface temperature only depends on $\zeta$ (spherical coordinate system ($r,\xi,\zeta$)).
		The sphere's radius is $r_0$. Gas particles impinge at temperature $T_\gas^{\ominus}$ and scatter at $T_\gas^{\+}$.
		}	
\end{figure}

Below, we describe a homogeneous solid particle by a sphere of radius $r_0$.
Non-spherical particles can be quantified in the same way \citep{ChrisDiss, Loesche2013} by using the radius of a volume-equivalent sphere, yielding an average force exerted on the particle.
Also inhomogeneous particles can be described by the same means as homogeneous spheres \citep{ChrisDiss, Loesche2013}.

Being subject to illumination from a fixed direction of incidence (e.g., $-\mathbf{e}_z$, see \fig{fig:sphere_with_Trot}), the surface temperature $T(r_0,\xi,\zeta)$ for a homogeneous sphere is only depending on the spherical coordinate $\zeta$ due to symmetry.

The model is subdivided into two parts. It comprises a kinetic model to describe the interaction of the gas with the particle surface in the free molecule regime and a heat transfer model to describe the particle heating due to irradiation, including interaction with the gas and thermal radiation.
The gas is assumed to be in thermal equilibrium within a large area around the suspended particle.

\subsection{Kinetic model}
For the kinetic description the gas molecule density $\sigma(\mathbf{r}, \mathbf{v}, t)$ with normalization $\int\sigma(\mathbf{r}, \mathbf{v}, t)\dd^3\mathbf{r}\dd^3\mathbf{v} = N$ (gas molecule count) is used.
The presence of a particle suspended in an effectively infinite gas imposes a boundary condition on $\sigma$, formally written as
\begin{equation}
	\sigma(\mathbf{r},\mathbf{v},t)\big|_{\partial V} =
	\begin{cases}
		\sigma^\-(\mathbf{r},\mathbf{v},t) & \mathbf{n}\cdot\mathbf{v}<0\\
		\sigma^\+(\mathbf{r},\mathbf{v},t) & \mathbf{n}\cdot\mathbf{v}>0 \; ,
	\end{cases}
	\label{eq:velocity_half-spaces}
\end{equation}
with $\mathbf{n}$ denoting the normal vector to the surface.
In the following, the marks `$\+$' and `$\-$' always restrict a physical variable to one of the two velocity half-spaces $\mathbf{n}\cdot\mathbf{v}>0$ and $\mathbf{n}\cdot\mathbf{v}<0$, respectively.
In other words, the index `$\-$' distinguishes the physical variables which are related to the undisturbed gas molecules from those molecules which have interacted with the particle, marked with the index `$\+$'.
Correspondingly, $T_\gas^\+$ and $T_\gas^\-$ denote the temperature of the gas molecules in their respective velocity half-spaces.
The subsequent balance of the momentum transfer between gas molecules and the suspended solid across its surface reads \citep{Hidy1970}
\begin{subequations}
	\begin{align}
		\mathbf{F} &= -\int\limits_{\partial V}\mathrm{d}\mathbf{A} \cdot \left( \boldsymbol{\underline{\Pi}}^\+ + \boldsymbol{\underline{\Pi}}^\- \right) \\
		\boldsymbol{\underline{\Pi}}^{\-/\+}(\mathbf{r},t) &= \int\limits_{\-/\+}\mathrm{d}^3v\, \sigma^{\-/\+}(\mathbf{r},\mathbf{v},t) \, m_\gas \, \mathbf{v} \otimes \mathbf{v} \; , \label{eq:gasForceB}
	\end{align}
	\label{eq:gasForce}
\end{subequations}
where $\boldsymbol{\underline{\Pi}}$ is the pressure or stress tensor.

We do not consider evaporation (\cite{Yalamov1976_photohoresis_co} did for the continuum regime), fragmentation and other processes, as we assume to remain below the melting temperature.
Furthermore, the gas does not penetrate the particle surface (first boundary condition)
\begin{equation}
	\mathbf{n}\cdot\left[ n^\+(\mathbf{r},t)\,\overline{\mathbf{v}^\+}(\mathbf{r},t) + n^\-(\mathbf{r},t)\,\overline{\mathbf{v}^\-}(\mathbf{r},t) \, \right] = 0 \; , \label{eq:mass_continuity_at_boundary}
\end{equation}
where the spatial gas density $n$ has been introduced (not to be confused with the normal vector $\mathbf{n}$).
$\overline{\mathbf{v}^{\+/\-}}$ denotes the component-wise averaged gas speed in the respective half-space.
Averages are defined by \eqref{eq:average}.
Therefore, for isotropic velocity distributions (a consequence of the thermal equilibrium of the gas), the net force exerted on a particle is (see \cite{ChrisDiss}, section 2.2 for details)
\begin{align}
   \mathbf{F} &= - \frac13\int\limits_{\partial V} m_\gas\,n\,\overline{\left(v^\-\right)^2}\dd\mathbf{A} \left( 1 + \left|\frac{\overline{v^\-_\mathbf{n}}}{\overline{v^\+_\mathbf{n}}}\right| \frac{\overline{\left(v^\+\right)^2}}{\overline{\left(v^\-\right)^2}} \right) \; , \label{eq:FphotIntegral_general}
\end{align}
and we simply write $n$ for $n^\-$ from now on, as it describes the unscattered gas.

Since the free molecular flow regime is characterized by large Knudsen Numbers $Kn$, it can be assumed that there are no collisions between gas molecules on the characteristic scale of the suspended body.
In this regime, for an isolated, force-free, thermally equilibrated, effectively infinite gas, the Boltzmann equation can be solved by the stationary and homogeneous dimensionless velocity distribution
\begin{equation}
   \sigma(\mathbf{v}) \sim \e^{-\frac{m_\gas \tilde{v}^2}{2\kB T_\gas}} \;.
\end{equation}
In the two velocity half-spaces $\+$ and $\-$, we therefore choose the two following Maxwell-Boltzmann-based velocity distributions with thermal and momentum accommodation (second boundary condition)
\begin{subequations}
	\begin{align}
	\sigma^\-(\mathbf{v}) &= n \, \sigma_0^\-(\mathbf{v}) = n \left(\frac{m_\gas}{2\pi \kB T_\gas^\-}\right)^{3/2} \e^{-\frac{m_\gas v^2}{2\kB T_\gas^\-}} \\
	\sigma^\+(\mathbf{v}) &= \alpha_\text{m} \, n^\+ \left(\frac{m_\gas}{2\pi \kB T_\gas^\+}\right)^{3/2} \e^{-\frac{m_\gas v^2}{2\kB T_\gas^\+}} + \left(1-\alpha_\text{m}\right)\,\sigma^\-(\mathbf{v}^\+_0) \\
	T_\gas^\+ &= T_\gas^\- + \alpha \left(T-T_\gas^\-\right) \label{eq:thermal_accommodation}\\
	\mathbf{v}^\+_0 &= \mathbf{v} - 2\mathbf{n}\left(\mathbf{n}\cdot\mathbf{v}\right) \; ,
	\end{align}
	\label{eq:newApproximation_sigma}
\end{subequations}
introducing the respective coefficients $\alpha$ and $\alpha_\text{m}$ being the thermal and momentum accommodation coefficient.
For thermal equilibrium between gas and surface, both coefficients reside in the interval [0,1], otherwise this is not necessarily the case \citep{Goodman1974alpha}.
Applying \eqref{eq:gasForce} together with the boundary conditions given by Eqs. \ref{eq:mass_continuity_at_boundary} and \ref{eq:newApproximation_sigma} yields the kinetic equation for the photophoretic force in the free molecule regime
\begin{align}
   \mathbf{F}_\text{phot} &= - \frac12\int\limits_{\partial V}\mathrm{d}\mathbf{A}\,\alpha_\text{m}\,p\left(1+\sqrt{\frac{T_\gas^\+}{T_\gas^\-}}\right) \; . \label{eq:integral_F}
\end{align}
This integral covers both $\Delta T$- and $\Delta\alpha_{(\text{m})}$ photophoresis, i.e. due to the variation of the surface temperature or the variation of the thermal or momentum accommodation coefficient across the surface.

In this paper we present a powerful approximation for $\Delta T$ photophoretic forces exerted on homogeneous spheres, resulting from directed illumination as shown in \fig{fig:sphere_with_Trot}.
Due to homogeneity, the sphere is assumed to have a rotational symmetric surface temperature, and the integral reduces to ($x=\cos\zeta$)
\begin{align}
	\mathbf{F}_\text{phot} &= - \pi\, r_0^2 \, p \,\int\limits_{-1}^{1} \alpha_\text{m}\sqrt{\frac{T_\gas^\+}{T_\gas^\-}}\,x \dd x \, \mathbf{e}_z \; , \label{eq:integral_F_sphere}
\end{align}
yielding the longitudinal photophoretic force.
Consequently $\alpha$ and $\alpha_\text{m}$ are considered constants further on.

\subsection{Surface temperature: Heat transfer problem}\label{sec:ht}
For the general case, where $T$ is unknown, it can be obtained by solving a heat transfer problem.
In a spherical system with rotational symmetry, the solution is basically a series of Legendre polynomials.
For convenience, the ansatz for the heat transfer problem is constructed insofar as the surface temperature is given by
\begin{align}
	T(r_0,\zeta) &= \sum\limits_{\nu=0}^{\infty} A_\nu \, P_\nu(\cos\zeta) \; . \label{eq:newApproximation_Tsurface}
\end{align}
Eventually, the linearization of $\sqrt{T_\gas^\+}$ in \eqref{eq:integral_F_sphere} to its first order at the mean temperature $\overline{T_\gas^\+}$ (see \eqref{eq:linearization}), and the application of Legendre Polynomial's orthogonality relation (see \eqref{eq:orthogonality}) collapses the integral to
\begin{align}
	\mathbf{F}_\text{phot} &\simeq -\frac{\pi}{3} \, \alpha \, \alpha_\text{m} \, \frac{p}{\sqrt{\overline{T_\gas^\+}\,T_\gas^\-}} \, r_0^2 \, A_1 \, \mathbf{e}_z \; , \label{eq:FphotBaseApprox1}
\end{align}
and the force is approximately only a function of $A_1$.
The mean temperature of the scattered gas is (with \eqref{eq:thermal_accommodation})
\begin{equation}
\overline{T_\gas^\+} = T_\gas^\-+\alpha\left(\overline{T}-T_\gas^\-\right) \; . \label{eq:mean_T_plus}
\end{equation}
The mean surface temperature $\overline{T}$ of the particle is solely determined by the 0-th expansion coefficient
\begin{equation}
\overline{T} = \frac1{4\pi}\int\limits_0^{2\pi}\int\limits_0^\pi T(\zeta) \sin\zeta \dd\zeta\dd\xi \stackrel{\text{\eqref{eq:newApproximation_Tsurface}}}{=} A_0 \; . \label{eq:mean_temperature} 
\end{equation}

\begin{figure}[ht!]
	\centering
	\includegraphics[width=\columnwidth]{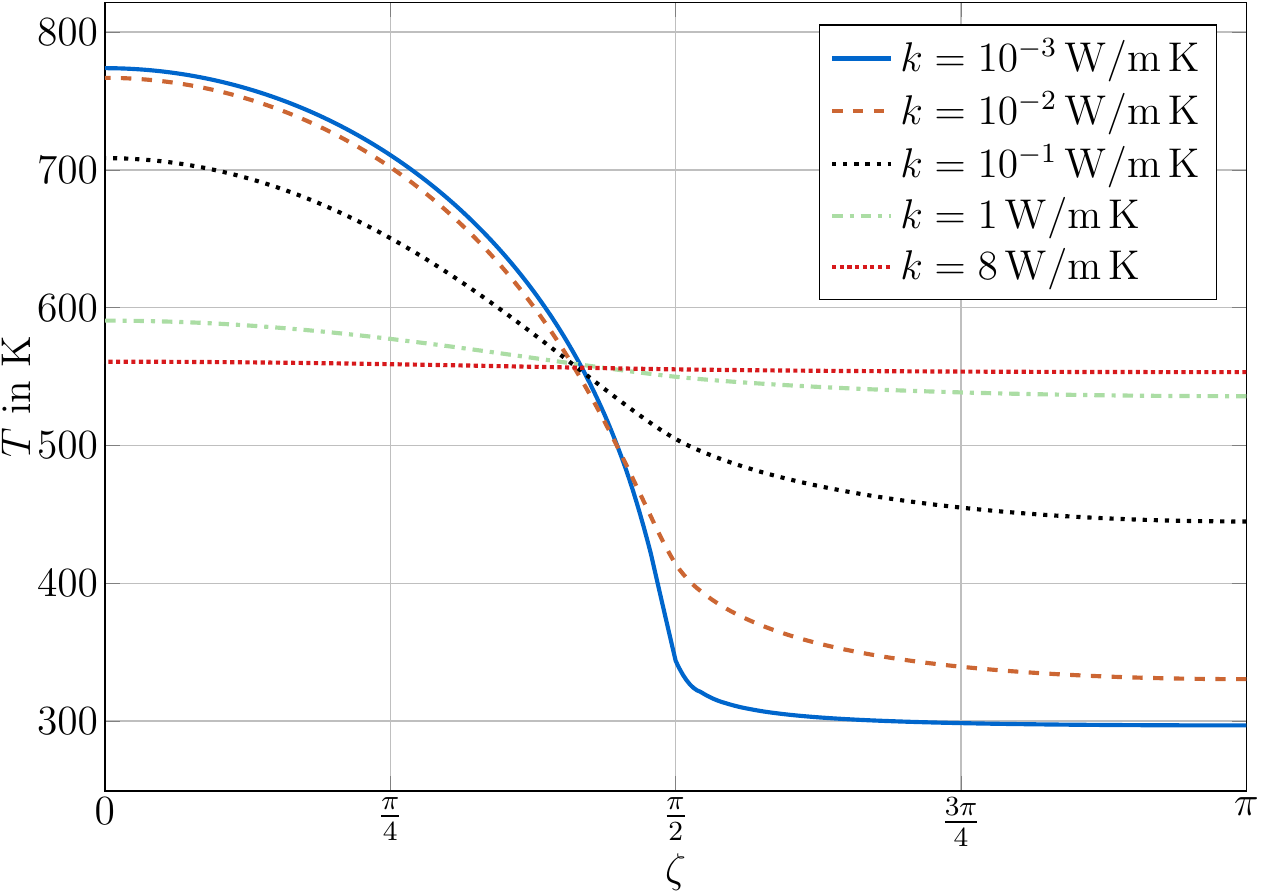}
	\caption{\label{fig:mantle_temperatures}
		Temperature distribution across the surface of a sphere with $r_0 = 0.66\u{mm}$ at different thermal conductivities, along the model setup shown in \fig{fig:sphere_with_Trot}. The intensity is $I_0=20\u{kW\,m^{-2}}$ ($\varepsilon=1$), the radiation temperature is $T{_\rad}=293\u{K}$ at $h=0\u{kW\,m^{-2}\,K^{-1}}$.
	}	
\end{figure}

We note that in \eqref{eq:FphotBaseApprox1}, the intrinsic linearization error of the square root $\sqrt{T_\gas^\+}$ in \eqref{eq:integral_F}/\eqref{eq:integral_F_sphere} is introduced, which basically all available approximations for longitudinal \textit{fm} photophoresis contain \citep{Hidy1967,Yalamov1976_photohoresis_fm,Beresnev1993,Rohatschek1995}. The deviations from the true value are more or less significant, depending on where $\sqrt{T_\gas^\+}$ was linearized at (except for \cite{Yalamov1976_photohoresis_fm}, the classical approximations use $T_\gas^\-$ due to their condition $\overline{T}\simeq T_\gas^{\-}$; for details see sec. \ref{sec:comparison}).
To avoid this error, \cite{Tong1973} suggested a numerical evaluation of the square root for higher order series of $T^\+$.

However, small particles, but also larger ones with a higher thermal conductivity will experience a more uniform heat-up, therefore the linearization of the square root suffices well. Conversely, if the temperature gradient is too strong, the introduced error can be significant.
We will discuss the solutions of the heat transfer problem in sec. \ref{sec:lösungsraum} and show that the numbers
\begin{align}
\varphi_\- &= \frac{I\,r_0}{k\,T_\gas^\-} \qquad\text{and} \qquad\varphi_\rad = \frac{I\,r_0}{k\,T_\rad} \; , \nonumber
\end{align}
but also $T_\rad$ and $T_\gas^\-$, respectively, are important measures for the quality of the linearization made in \eqref{eq:FphotBaseApprox1}.
As an example, in \fig{fig:mantle_temperatures} we show exemplary mantle temperatures for a sphere of $r_0 = 0.66\u{mm}$, experiencing directed illumination of $I_0=20\u{kW\,m^{-2}}$ ($\varepsilon=1$) at $T_\rad=293\u{K}$ with $h=0\u{kW\,m^{-2}\,K^{-1}}$. For $k<0.1\u{W\,m^{-1}\,K^{-1}}$ the linearization of $\sqrt{T_\gas^\+}$ at $\overline{T_\gas^\+}$ will introduce errors, as the temperature curves lack symmetry.

\subsubsection{Ansatz}
The governing equation for the heat transfer problem we consider has the form ($I=\varepsilon\,I_0$ is the reduced intensity)
\begin{equation}
   k\,\Delta T = - I\, q(r,\cos\zeta) \; , \label{eq:heatEQ}
\end{equation}
where $q$ is the normalized heat source function with
\begin{equation}
   \int\limits_V q(r,\zeta)\dd V = \pi\, r_0^2 \; . \label{eq:q_integral}
\end{equation}
For the general solution
\begin{subequations}
\begin{equation}
   T(r,\zeta) = T_1(r,\zeta)+T_2(r,\zeta)
\end{equation}
the homogeneous and particular ansatz functions, respectively, are
\begin{align}
	T_1(r,\zeta) &= \sum\limits_{\nu=0}^{\infty}\left(A_\nu-B_\nu\,J_\nu(r_0)\right)\, \left(\frac{r}{r_0}\right)^\nu P_\nu(\cos\zeta) \label{eq:T_h} \\
	T_2(r,\zeta) &= \sum\limits_{\nu=0}^{\infty}B_\nu\,J_\nu(r) \,P_\nu(\cos\zeta) \; . \label{eq:T_i}
\end{align}
\label{eq:general_solution_temp}
\end{subequations}
Then, by construction, $T_1(r_0,\zeta)+T_2(r_0,\zeta)$ yield the surface temperature as \eqref{eq:newApproximation_Tsurface}.
The particular solution employs the asymmetry factor $J_\nu$
\begin{subequations}
	\begin{align}
		J_\nu(r) &= \frac1{r_0} \left[
		r^{-\nu-1} \int\limits_{0}^{r} s^{\nu+2} q_\nu(s) \dd s + 
		r^\nu \int\limits_{r}^{r_0} s^{\nu-1} q_\nu(s) \dd s 
		\right] \\
		q_\nu(r) &= \frac{2 \nu+1}{2} \int\limits_{-1}^{1}q(r,x)\,P_\nu(x)\dd x \label{eq:q_Legendre_expansion_coefficients} \\
		J_\nu &\equiv J_\nu(r_0) = \int\limits_{0}^{r_0} \left(\frac{r}{r_0}\right)^{\nu+2} q_\nu(r) \dd r \; . \label{eq:asymmetry_factor_J}
	\end{align}
	\label{eq:asymmetry_factor_all}
\end{subequations}
$q_\nu(r)$ are the Legendre expansion coefficients of the source $q$.
This construction gets clearer upon applying the Laplace operator on the particular solution $T_2$, which yields the Legendre series of the inhomogeneity $q$
\begin{align}
	-\frac{k}{I}\,\Delta T_2(r,\zeta) &= \sum\limits_{\nu=0}^{\infty} q_\nu(r)\,P_\nu(\cos\zeta) \equiv q(r,\cos\zeta)\; ,
\end{align}
for the right coefficients $B_\nu$ (see sec. \ref{sec:solution}). This eventually reproduces the heat equation (\eqref{eq:heatEQ}).

\subsubsection{Asymmetry factor}
Obtaining $J_1$ is generally complicated. Several  works \citep{Yalamov1976_photohoresis_fm,Yalamov1976_photohoresis_co,Dusel1979,Arnold1984,Greene1985,Mackowski1989aerosol,Xu1999,Ou2005_photophoresis,Li2010absorption} determined it, e.g., for Mie scattering (usage of corresponding normalized source function $q$). For perfectly absorbing spherical particles, the entire radiation is deposited at the surface, which is often used.
The corresponding normalized source for light towards the direction $-\mathbf{e}_z$ reads (\fig{fig:sphere_with_Trot}, i.e. pointing downwards)
\begin{equation}
   q(r,\zeta) = \delta\left(r-r_0\right)\, \Theta\left(\pi/2-\zeta\right)\,|\cos\zeta| \; ,
\end{equation}
and $J_0$ and $J_1$ yield (\eqref{eq:asymmetry_factor_J})
\begin{subequations}
\begin{align}
   J_0 &= \frac14 \label{eq:J0} \\
   J_1 &= \frac12 \, . \label{eq:J1}
\end{align}
\end{subequations}
For light into the direction $\mathbf{e}_z$ it is $q(r,\zeta) = \delta\left(r-r_0\right)\, \Theta\left(\zeta-\pi/2\right)\,|\cos\zeta|$, and therefore $J_1 = -1/2$.
The factor $J_1= 1/2$ is commonly used in \citet{Wurm2010,Rohatschek1995} and others, as generally the positive $z$-axis is assigned to the illuminated half-sphere.
Along \eqref{eq:q_integral}, the deposited power is given by $4\pi r_0^2\,I\,J_0$.

\subsubsection{Boundary conditions}
To account for thermal radiation, the following boundary condition was chosen
\begin{align}
   -\left.k\frac{\partial T}{\partial \mathbf{n}}\right|_{\partial V} &= h\left(T-T_\gas^{\-}\right) + \sigma_\text{SB}\varepsilon\left(T^4-T_\rad^4\right) \; . \label{eq:boundary_condition_HT}
\end{align}
To prevent nonlinear mixing of the expansion coefficients $A_\nu$ and $B_\nu$ in \eqref{eq:general_solution_temp} at multiple orders, the term $\sigma_\text{SB}\varepsilon(T^4-T_\rad^4)$ will be linearized at the yet unknown mean temperature $\tilde{T}$
\begin{align}
   \sigma_\text{SB}\varepsilon(T^4-T_\rad^4) &= \sigma_\text{SB}\varepsilon\left(4 T\, \tilde{T}^3-T_\rad^4 -3 \tilde{T}^4\right) + \dots \\
   \tilde{T} &= \left( \frac1{4\pi}\int\limits_0^{2\pi}\int\limits_0^\pi T(\zeta)^4 \sin\zeta \dd\zeta\dd\xi \right)^{1/4} \; . \label{eq:mean_temperature4}
\end{align}
The first addend in \eqref{eq:boundary_condition_HT} accounts for the energy flux at the surface between particle and gas, which is (for the gas molecule density distributions in \eqref{eq:newApproximation_sigma})
\begin{equation}
   \mathbf{n}\cdot\left[ \int\limits_{\+}\mathrm{d}^3v\, \sigma^\+ \, \mathbf{v} \, \frac12 m_\gas \, v^2+\int\limits_{\-}\mathrm{d}^3v\, \sigma^\- \, \mathbf{v} \, \frac12 m_\gas \, v^2\right] = h\left(T-T_\gas^{\-}\right) \; ,
\end{equation}
introducing the heat transfer coefficient in this context as ($\overline{v}$ is the mean of the gas molecule speed $v=\|\mathbf{v}\|$ along \eqref{eq:average})
\begin{align}
   h &= \alpha_\text{m} \, \alpha \, p \, \sqrt{\frac{2\kB}{\pi\,T_\gas^\-\,m_\gas}} = \frac12 \alpha_\text{m} \, \alpha \frac{p}{T_\gas^\-}\overline{v} \; . \label{eq:h}
\end{align}
For diatomic gases the factor $\frac12$ in $h$ turns to $\frac34$ \citep{Rohatschek1985}.

\subsubsection{Solution}\label{sec:solution}
The expansion coefficients $B_\nu$ in the general solution \eqref{eq:general_solution_temp}
\begin{equation}
T(r,\zeta) = \sum\limits_{\nu=0}^{\infty}\left(A_\nu-B_\nu\,J_\nu(r_0)\right)\, \left(\frac{r}{r_0}\right)^\nu P_\nu(\cos\zeta) + \sum\limits_{\nu=0}^{\infty}B_\nu\,J_\nu(r) \,P_\nu(\cos\zeta) \nonumber
\end{equation}
can be obtained by inserting \eqref{eq:T_i} in \eqref{eq:heatEQ} as
\begin{equation}
   B_\nu = \frac{I \,r_0}{(2\nu+1)k} \label{eq:koeff_B} \; .
\end{equation}
Conversely, the coefficients $A_\nu$ are determined by integrating \eqref{eq:boundary_condition_HT} with $P_\nu$ in [-1,1] together with the identity
\begin{equation}
   J'_\nu(r_0) = -\frac{1+\nu}{r_0}J_\nu(r_0) \; ,
\end{equation}
yielding $A_\nu$ as
\begin{subequations}
	\begin{align}
	A_\nu &= \frac{I\,J_\nu}{\nu\frac{k}{r_0} + h + 4\sigma_\text{SB}\varepsilon\,\tilde{T}^3} \qquad \nu\ge 1 \label{eq:Ai_fm}\\
	A_0 &= \frac{h\,T_\gas^\- + \sigma_\text{SB}\varepsilon\,\left(3\tilde{T}^4+T_\rad^4\right)+I\,J_0}{h + 4\sigma_\text{SB}\varepsilon\,\tilde{T}^3} \stackrel{\eqref{eq:mean_temperature}}{=} \overline{T} \; . \label{eq:A0_fm}
	\end{align}
	\label{eq:newApproximation_coefficients}
\end{subequations}

The unknown temperatures $\overline{T}$ and $\tilde{T}$ can be related by integrating the boundary condition \eqref{eq:boundary_condition_HT} around the sphere, and using Gauss's theorem
\begin{align}
-k\int\limits_{\partial V} \boldsymbol{\nabla}T\cdot\mathrm{d}\mathbf{A} &= \int\limits_{\partial V} \left( h\left(T-T_\gas^{\-}\right) + \sigma_\text{SB}\varepsilon\left(T^4-T_\rad^4\right) \right)\dd A \nonumber \\
&= -k\int\limits_V \Delta T \dd V \nonumber \\
&\stackrel{\text{\eqref{eq:heatEQ}}}{=} \varepsilon\, I_0 \int\limits_V q(r,\zeta)\dd V \stackrel{\text{\eqref{eq:q_integral}}}{=} \pi r_0^2 \,\varepsilon\, I_0 \; .
\end{align}
Then, the two temperatures $\overline{T}$ and $\tilde{T}$ (Eqs. \ref{eq:mean_temperature} and \ref{eq:mean_temperature4}) meet the balance (mean value theorem)
\begin{equation}
\pi r_0^2\, \varepsilon \, I_0 = 4\pi r_0^2\left( h\left(\overline{T}-T_\gas^{\-}\right) + \sigma_\text{SB}\varepsilon\left(\tilde{T}^4-T_\rad^4\right) \right) \; . \label{eq:mean_temperatures_balance}
\end{equation}

For $h\ll 4\sigma_\text{SB}\varepsilon\,\tilde{T}^3$ the particle is basically in radiative equilibrium along the boundary condition \eqref{eq:boundary_condition_HT}, therefore $\tilde{T}$ is equivalent to the black-body temperature 
\begin{equation}
   \lim\limits_{h\to 0} \tilde{T} \stackrel{\eqref{eq:mean_temperatures_balance}}{=} T_\bb \; ,
\end{equation}
given as
\begin{equation}
	T_\bb = \sqrt[4]{\frac{I_0}{4\sigma_{\text{SB}}}+T_\rad^4} \label{eq:blackBodyTemp}  \; .
\end{equation}
Similarly, with \eqref{eq:J0} it can be inferred, that 
\begin{equation}
	\lim\limits_{h\to 0} \overline{T} \stackrel{\text{\eqref{eq:A0_fm}}}{=} T_\bb \; .
\end{equation}

This coincidence is a direct result of the linearization of the boundary condition \eqref{eq:boundary_condition_HT}.
Generally, for functions $f(x)$, unless $f\simeq\text{const}$, both mean values are not similar $\overline{|f|} \ncong \overline{|f|^4}^{1/4}$, as the first and fourth norm are not equal.
This can also be understood as for the integration of $|f|^4$ the coefficients of the Legendre polynomial expansion of $|f|$ mix up and multiple coefficients contribute, not solely $A_0$.
Though for small particles --- or large particles that are good conductors ---, the 0-th temperature expansion coefficient will dominate $A_0\gg A_1$ ($T\simeq\text{const}$), and therefore both means are comparable $\tilde{T} \approx \overline{T}$.
Large particles with no high enough $k$ generally will develop a stronger temperature gradient during direct illumination, just as small particles do for low thermal conductivities $k$ (\fig{fig:sphere_with_Trot}), therefore the ratio $r_0/k$ in $\varphi_\rad$ plays a role here.
For example, numerical calculations on two spheres of $r_0=1\u{m}$ and $r_0=1\u{mm}$ with $k=0.1\u{W\,m^{-1}\,K^{-1}}$ and $h=0\u{kW\,m^{-2}\,K^{-1}}$ yielded mean temperatures of $\overline{T}=462.3\u{K}$ and $\overline{T}=551.0\u{K}$, respectively, which are not the black-body temperature of $T_\bb=556.0\u{K}$ at $I=20\u{kW\,m^{-2}}$ and $T_\rad=293.2\u{K}$.

Conversely, for the cases where $h$ contributes, the unknown mean temperatures $\overline{T}$ and $\tilde{T}$ can be determine by solving \eqref{eq:A0_fm} and \ref{eq:mean_temperatures_balance}.
Because of contact with the cooler gas it can be expected that $\tilde{T}$ and $\overline{T}=A_0$ will be lower than in the case where $h\to 0$.

\subsection{Result}\label{eq:result}
In \eqref{eq:FphotBaseApprox1}, only the coefficient $A_1$ contributes to the photophoretic force \footnote{reminding, that the square root in \eqref{eq:integral_F} was linearized at $\overline{T_\gas^\+}$}.
Inserting the previously found $A_1$ (\eqref{eq:Ai_fm}, $J_0=1/4$, as \eqref{eq:q_integral} holds) yields
\begin{subequations}
\begin{equation}
   \mathbf{F}_\text{phot} \simeq-\frac{\pi}{3} \, \alpha \, \alpha_\text{m} \frac{p}{\sqrt{\overline{T_\gas^\+}\,T_\gas^\-}} \, r_0^2 \, 	\frac{I\,J_1}{\frac{k}{r_0}+h+4\sigma_\text{SB}\varepsilon\,\tilde{T}^3} \, \mathbf{e}_z \label{eq:Fphot}
\end{equation}
with
\begin{align}
		\tilde{T} &= \frac{1}{2} \left(\sqrt{\frac{\sqrt{2} h}{\sigma_\text{SB}  \sqrt{\psi} \varepsilon }-2 \psi}-\sqrt{2\psi}\right) \label{eq:T_tilde} \\
		\overline{T} &= \frac{h\,T_\gas^\- + \sigma_\text{SB}\varepsilon\,\left(3\tilde{T}^4+T_\rad^4\right)+I/4}{h + 4\sigma_\text{SB}\varepsilon\,\tilde{T}^3} \label{eq:T_bar} \\
		\overline{T_\gas^\+} &= T_\gas^\-+\alpha\left(\overline{T}-T_\gas^\-\right)
	\end{align}
	\label{eq:f_fm}
\end{subequations}
and the auxiliary variables
\begin{subequations}
	\begin{align}
		\kappa &= \sigma_\text{SB}\varepsilon \left(\sqrt{81 h^4 + 12 \sigma_\text{SB}  \varepsilon  \left(4 h T_\gas^\-+\varepsilon  \left(I_0+4 \sigma_\text{SB}  T_\rad^4\right)\right)^3}+9 h^2\right) \label{eq:kappa} \\
		\psi &= \frac{\frac{\sqrt[3]{2}}{\sigma_\text{SB}\varepsilon} \kappa^{2/3}-8 \sqrt[3]{3} h T_\gas^\--2 \sqrt[3]{3} \varepsilon \left(I_0+4 \sigma_\text{SB}  T_\rad^4\right)}{2\ 6^{2/3} \sqrt[3]{\kappa}} \; . \label{eq:psi}
	\end{align}
	\label{eq:kappa_and_psi}
\end{subequations}
\eqref{eq:kappa} and \ref{eq:psi} arise from solving the quartic equation for $\tilde{T}$ (\eqref{eq:T_tilde}), which forms when \eqref{eq:A0_fm} is inserted into \ref{eq:mean_temperatures_balance}.
$\tilde{T}$ (\eqref{eq:T_tilde}) is very close to the numerical values (solving \eqref{eq:heatEQ} with the non-linear boundary condition \eqref{eq:boundary_condition_HT}, also see \ref{sec:comparison}; or solving the balance \eqref{eq:mean_temperatures_balance} with the assumption $\overline{T}=\tilde{T}$).
Conversely, \eqref{eq:T_bar} to determine $\overline{T}$ is independent from $r_0$ and $k$, and thus it deviates from the real values as $k/r_0$ gets smaller.
But this does only have little consequences for $\mathbf{F}_\text{phot}$ (\eqref{eq:Fphot}) though (as the mantle temperature will become non-symmetric anyway, as already discussed in sec. \ref{sec:ht}), which will be shown in sec. \ref{sec:comparison}.

For $T_\rad=T_\gas^\-$ and low intensities $I$, i.e. $\lim\limits_{I\to 0}\tilde{T} \approx T_\rad$, \eqref{eq:Fphot} turns into \cite{Beresnev1993}'s Eq. 28, which is the free molecule regime limit for the far more advanced kinetic model in \cite{Beresnev1993}, which intentionally also covers parts of the transition regime. 

For $h\ll 4\sigma_\text{SB}\varepsilon\,\tilde{T}^3$, above equations reduce to
\begin{subequations}
	\begin{align}
	\mathbf{F}_\text{phot} &\simeq-\frac{\pi}{3} \, \alpha \, \alpha_\text{m} \frac{p}{\sqrt{\overline{T_\gas^\+}\,T_\gas^\-}} \, r_0^2 \, 	\frac{I\,J_1}{\frac{k}{r_0}+4\sigma_\text{SB}\varepsilon\,T_\bb^3} \, \mathbf{e}_z \label{eq:Fphot2}\\
	\overline{T_\gas^\+} &= T_\gas^\-+\alpha\left(T_\bb-T_\gas^\-\right) \\
	T_\bb &= \sqrt[4]{\frac{I_0}{4\sigma_{\text{SB}}}+T_\rad^4} \; .
	\end{align}
	\label{eq:f_fm2}
\end{subequations}
This is especially the case for low pressures.

\subsection{Comparison to standard approximations}\label{sec:comparison}
It is of interest to know the behavior of present approximations for particles in radiative equilibrium with an external radiation field and being notably hotter/colder than the surrounding gas, i.e. $|\overline{T}/T_\gas^\-| \ge 1$ or  $|\overline{T}/T_\gas^\-| \le 1$. 
Basically three different standard approximations, which are valid for $|\overline{T}/T_\gas^\-| \simeq 1$ --- i.e. the particle's mean surface temperature basically being the gas temperature --- are examined for this new setting to justify the need for our new approximation.
Cases with the particle on average being much hotter or colder than the surrounding gas were mentioned earlier.
	
\cite{Yalamov1976_photohoresis_fm,Hidy1967,Rohatschek1995,Tong1973} basically use a kinetic model that also employs Max-Boltzmann velocity distributions --- valid for the \textit{fm} regime ---, therefore the force as a function of the particle surface temperature $T$ can be written as \eqref{eq:integral_F} (but only \cite{Yalamov1976_photohoresis_fm} incorporates momentum accommodation, therefore $\alpha_\text{m}$ is not present in the other publications).
	
\cite{Beresnev1993,Chernyak1993} use an advanced kinetic model with momentum accommodation (normal and tangential), enabling their results not only to be valid for the free molecule regime but also to cover parts of the transition regime. Nonetheless, they provide an equation for the \textit{fm}-regime limit, which we discuss here.

All approximations are very similar in structure to \eqref{eq:f_fm}, with $\tilde{T}\to 0$ for \citet{Yalamov1976_photohoresis_fm,Hidy1967,Rohatschek1995,Tong1973}, and $\lim\limits_{I\to 0}\tilde{T} \to T_\rad$ and $T_\rad=T_\gas^\-$ for \cite{Beresnev1993}. Except for \cite{Yalamov1976_photohoresis_fm}, all models use $\overline{T_\gas^\+}\to T_\gas^\-$.
\cite{Tong1973} does not provide an approximation for the integral equation \ref{eq:integral_F}, but suggests its numerical evaluation.

\begin{figure}[htbp]
	\includegraphics[width=\columnwidth]{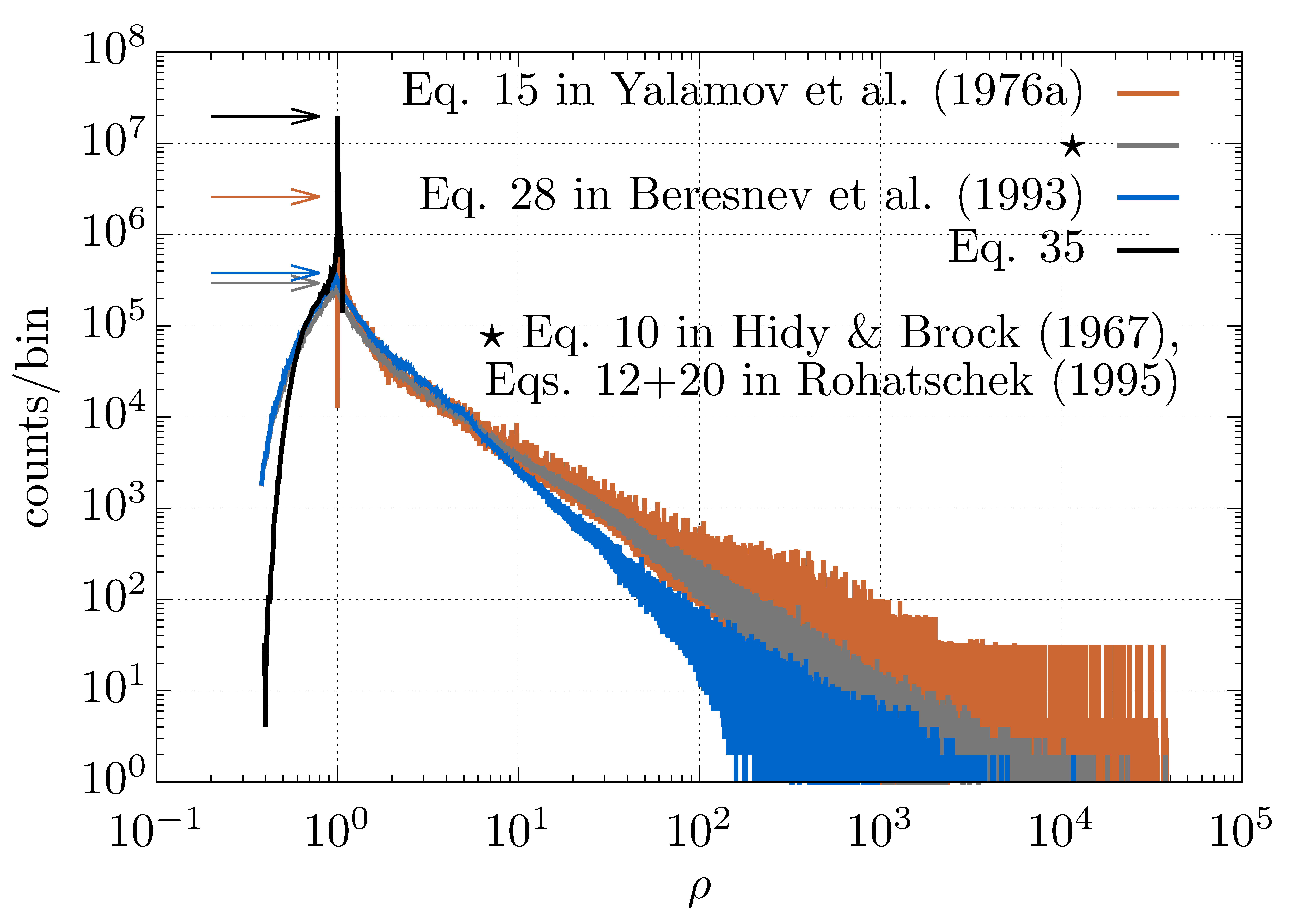}
	\caption{
		Parameter sweep histogram for $63\cdot 10^6$ parameter combinations.
		The parameter sweep intervals are given in \tab{tab:parameter_config}.
		$\rho$ (\eqref{eq:ratio}) is the ratio of the corresponding approximation of the photophoretic force and the total of the effectively exact numerical result given by \eqref{eq:integral_F}, where the surface temperatures were numerically obtained with COMSOL.
		The bin size is 0.005 (0.5\%).
		Color-coded arrows point towards the respective histogram's peak.
		The histogram of the new approximation (black) is restricted to $0.40\le\rho\le1.07$ while the other approximations overestimate the force up to several orders of magnitude.
		}
	\label{fig:quality}
\end{figure}
For all subsequent considerations it is assumed that $h=0$.
In \tab{tab:statistics}, the quality of the new approximation compared to the standard approximations is discussed by their ratios with the total of \eqref{eq:integral_F}, which is our underlying reference:
\begin{equation}
	\rho = \frac{\text{approximation}}{|\mathbf{F}_\text{phot}|} \qquad\text{where }\mathbf{F}_\text{phot}\text{ is given by \eqref{eq:integral_F}} \; . \label{eq:ratio}
\end{equation}
As true value we denote the numerical integration of this integral based on the solution of the heat transfer problem, where the heat equation \eqref{eq:heatEQ} with the non-linear boundary condition \eqref{eq:boundary_condition_HT} was solved.
We do this by employing COMSOL for a parameter sweep of $63\cdot 10^6$ parameter combinations to obtain the necessary temperature distributions on the particle surface (details can be found in
\cite{Loesche2012,Loesche2013,ChrisDiss}).
The parameters range in the intervals given in \tab{tab:parameter_config}.
$\alpha_\text{m}=1$, as not all examined models incorporate momentum accommodation.
The accompanying histogram is shown in \fig{fig:quality}.
\begin{table*}[!htbp]
\centering
\caption{
	Statistical properties of $\rho$ for selected approximations for the photophoretic force arising from directed illumination.
	A parameter sweep of $63\cdot 10^6$ parameter combinations was performed along the parameter intervals given in \tab{tab:parameter_config}.
	Values printed gray in round brackets are for $r_0$ restricted to the interval $[0.11,11]\u{mm}$.
	 }
\label{tab:statistics}
\begin{tabular}{$C{5cm} >{\bfseries}^C{1cm}>{\bfseries}^C{1cm} ^C{1cm}^C{1cm}>{\itshape}^C{1cm}}
\toprule
approximations for longitudinal \textit{fm}-photophoresis & min & max & mean & median & STD \\
\midrule
Eq. 10 in \cite{Hidy1967}, 12+20 in \cite{Rohatschek1995} 	& 0.38 \textcolor{gray}{(0.38)}	& 275\,022 \textcolor{gray}{(3\,037)} & 50.0 \textcolor{gray}{(7.78)}	& 2.33 \textcolor{gray}{(1.4)}	& 473.6 \textcolor{gray}{(27.43)} \\ \arrayrulecolor{lightgray}\hline
Eq. 15 in \cite{Yalamov1976_photohoresis_fm}				& 1.00 \textcolor{gray}{(1.00)} 	& 38\,318 \textcolor{gray}{(423.0)} & 45.1 \textcolor{gray}{(7.04)} 	& 2.25 \textcolor{gray}{(1.35)}	& 341.8 \textcolor{gray}{(19.33)} \\ \hline
Eq. 28 in \cite{Beresnev1993} 								& 0.38 \textcolor{gray}{(0.38)} 	& 108\,088 \textcolor{gray}{(1\,587)} & 8.29 \textcolor{gray}{(3.19)} 	& 1.45 \textcolor{gray}{(1.18)}	& 96.52 \textcolor{gray}{(10.10)} \\ \hline
\rowstyle{\bfseries} \eqref{eq:f_fm2}						& 0.40 \textcolor{gray}{(0.53)}	& 1.07 \textcolor{gray}{(1.07)} & 0.97 \textcolor{gray}{(0.99)} 	& 1.00 \textcolor{gray}{(1.00)}	& 0.10 \textcolor{gray}{(0.06)} 	\\ \arrayrulecolor{black}\bottomrule
\end{tabular}
\end{table*}
\begin{table}[h]
	\centering
	\caption{\label{tab:parameter_config}
		Intervals for the parameter sweep for the subsequent comparison of the standard approximations and the new approximation with the numerical values given from \eqref{eq:integral_F}  ($[a,b]$ denotes an interval between the numbers $a$ and $b$).
		$T_\gas^\ominus$ ranges from 50 to 1500 K in steps of 50 K, including the two values 10 K and 273 K.
		All intervals are equally subdivided (log scale; the additional `1 m' for $r_0$ means, there is a gap between 1 m and 0.11 m concerning this equal subdivision).		
		Details on the subdivision can be found in \cite{ChrisDiss}.
		}
	\begin{tabular}{c L{5.5cm}}
		\toprule
		parameter & parameter sweep intervals \\ \midrule
		$r_0$ & $[1.1\times 10^{-4}, 1.1\times 10^{-1}]\u{m}$, and 1 m\\
		$k$ & $[10^{-3}, 8]\u{W~m^{-1}~K^{-1}}$\\
		$\alpha$ & $[0.1, 1]$ \\
		$\alpha_\text{m}$ & 1\\
		$I$ & $[0.5, 40]\u{kW~m^{-2}}$\\
		$T_\gas^\ominus$ & $[10, 1500]\u{K}$ \\
		$T_\rad$ & $[0, 350]\u{K}$ \\ \bottomrule
	\end{tabular}
\end{table}

\fig{fig:quality} and \tab{tab:statistics} show that the approximation given by \eqref{eq:f_fm2} is generally more accurate for longitudinal photophoresis for $|\overline{T}/T_\gas^\-| \ge 1$ or  $|\overline{T}/T_\gas^\-| \le 1$.
Other approximations tend to overestimate the photophoretic force up to several orders of magnitude, since their validity is given only for small intensities, so that the particle's mean surface temperature approximately corresponds to the gas temperature $T_\gas^\-$.
Comparing the values in \tab{tab:statistics} for $r_0$ up to 1 m (printed black) and up to 11 mm only (printed gray) suggests, that the new approximation's relative error interval basically remains constant, as the other approximations' error interval drastically grows with particle size $r_0$. 
We emphasize here, that the varied parameters do not represent statistical data, and therefore \fig{fig:quality} and \tab{tab:statistics} have no strict mathematical meaning but rather show the behavior and reliability of all approximations and provide a relative error interval based on the chosen parameter range in this paper.

The new approximation \eqref{eq:f_fm2} accounts for thermal radiation and therefore also supports particle temperatures that significantly deviate from the gas temperature. For the investigated parameters, the maximum underestimation is about 50-60\%, whilst the maximum overestimation is only 7\%. The best classic approximation is given by \cite{Yalamov1976_photohoresis_fm}, where the maximum overestimation is a factor of $38\,000$ (423 for $r_0$ only up to $11\u{mm}$) for the same parameters.

\subsection{Discussion}\label{sec:lösungsraum}
In this subsection, we discuss the influence of all parameters from which the new approximation depends on.
We basically concentrate on the case $h\ll 4\sigma_\text{SB}\varepsilon\,\tilde{T}^3$, where $h$ effectively does not contribute to the heat transfer problem.
This is also due to the evaluation of all approximations discussed in the previous section.
For a proper usage of the new approximation, we provide a condition for its perfect accuracy.

In our model, the heat transfer problem's solution depends on $r_0,\,k,\,I_0,\,\varepsilon,\,T_\rad$ and $T_\gas^\-$  (sec. \ref{sec:solution}), the photophoretic force additionally depends on $\alpha$ and $p$.
To obtain more information about the solutions, we define the units free variables
\begin{subequations}
	\begin{align}
	\lambda &= \frac{r}{r_0}\\
	\tau_\- &= \frac{T}{T_\gas^\-} \qquad\text{and} \qquad\tau_\rad = \frac{T}{T_\rad} \\
	\varphi_\- &= \frac{\varepsilon\,I_0\,r_0}{k\,T_\gas^\-} \qquad\text{and} \qquad\varphi_\rad = \frac{\varepsilon\,I_0\,r_0}{k\,T_\rad} \label{eq:phi} \\
	\vartheta_\rad &= \sigma_\text{SB}\frac{T_\rad^4}{I_0} \label{eq:theta} \; .
	\end{align}
	\label{eq:unitsFreeVars}
\end{subequations}
Rewriting the heat equation \eqref{eq:heatEQ} in this units free notation ($\tau$ and $\varphi$ are short for either $\tau_\-$ or $\tau_\rad$ and $\varphi_\-$ or $\varphi_\rad$, respectively) yields
\begin{align}
	\tilde{\Delta}\tau &= -\varphi\,\tilde{q}(\lambda,\cos\zeta) \; . \label{eq:heatEQ2}
\end{align}
$\tilde{\Delta}$ and $\tilde{q}$ can be found in sec. \ref{eq:laplace}. The boundary condition \eqref{eq:boundary_condition_HT} turns into
\begin{subequations}
	\begin{align}
	-\left.\frac{\partial\tau_\-}{\partial \mathbf{n}}\right|_{\partial V} &= \varphi_\- T_\gas^\-\frac{h}{I}\left(\tau_\--1\right) + \varphi_\- \frac{\sigma_\text{SB}\,\varepsilon}{I}\left(\left(\tau_\-T_\gas^\-\right)^4-T_\rad^4\right) \\
	\text{or}\nonumber\\
	-\left.\frac{\partial\tau_\rad}{\partial \mathbf{n}}\right|_{\partial V} &= \varphi_\rad \frac{h}{I}\left(\tau_\rad\,T_\rad-T_\gas^{\-}\right) + \varphi_\rad\,\vartheta_\rad\left(\tau_\rad^4-1\right) \; ,
	\end{align}
	\label{eq:boundary_condition_HT2}
\end{subequations}
respectively.
\begin{figure}[!h]
	\includegraphics[trim=10 35 225 20, clip, width=\columnwidth]{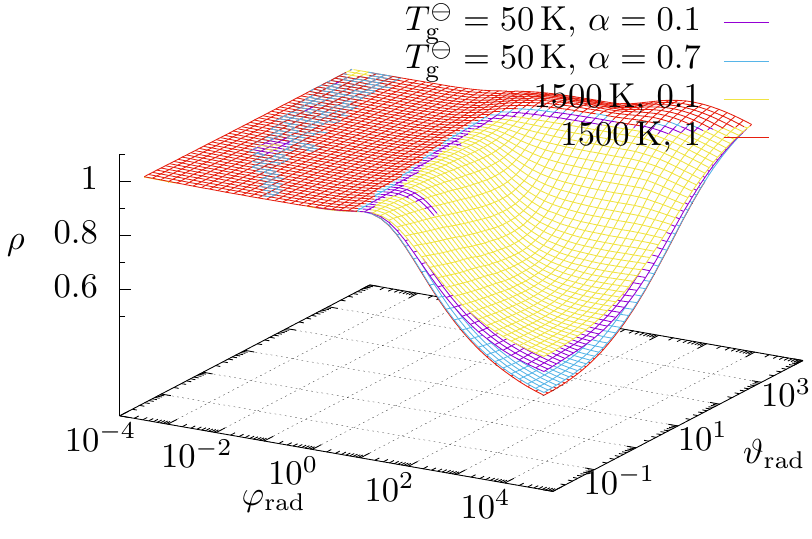}
	\caption{
		Here, $\rho(\varphi_\rad,\vartheta_\rad,\alpha, T_\gas^{\ominus})$ (\eqref{eq:ratio}) is the ratio of the new approximation \eqref{eq:f_fm2} and the total of the effectively exact numerical result given by \eqref{eq:integral_F}, where the surface temperatures were numerically obtained with COMSOL.
		Clearly visibly is a plateau, where the ratio is constantly 1, quasi-independently of $T_\gas^{\ominus}$ and $\alpha$.
		To examine this area, $\rho$ is replotted for $\alpha=1$ as contours in \fig{fig:error2}.
	}
	\label{fig:error}
\end{figure}
\begin{figure}[!h]
	\includegraphics[trim=5 25 185 10, clip, width=\columnwidth]{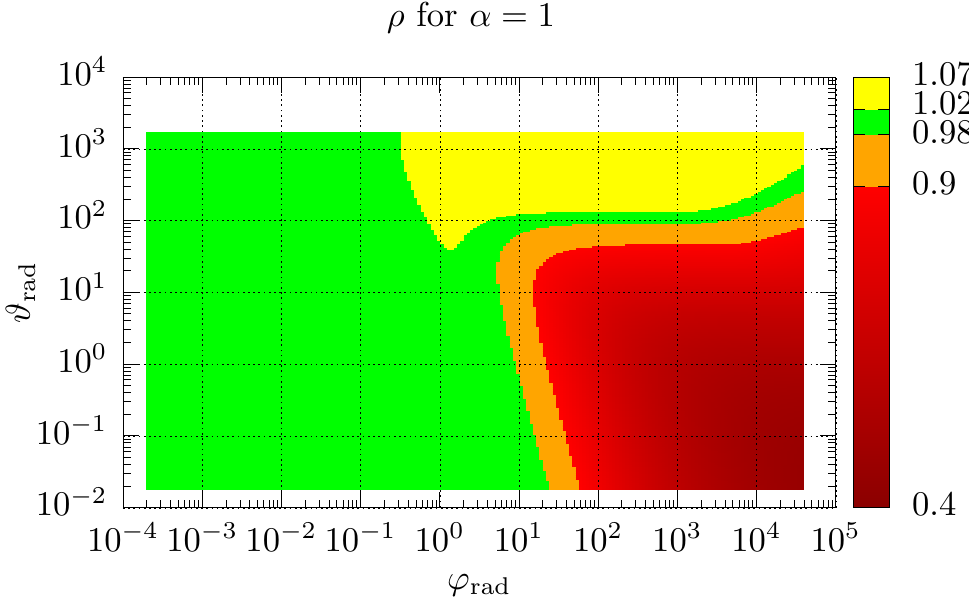}
	\caption{
		Here, $\rho(\varphi_\rad,\vartheta_\rad,\alpha, T_\gas^{\ominus})$ (\eqref{eq:ratio}) is the ratio of the new approximation \eqref{eq:f_fm2} and the total of the effectively exact numerical result given by \eqref{eq:integral_F}, where the surface temperatures were numerically obtained with COMSOL.
		Different ratio intervals are color-coded.
		The small interval $0.98\le\rho\le1.02$ is marked in green, where the ratio --- except for the areas close to the borders --- is constantly 1, quasi-independently of $T_\gas^{\ominus}$ and $\alpha$.
		In this green area, the relative error is $\le2\%$.
	}
	\label{fig:error2}
\end{figure}
	
For a vanishing $h$ ($h\ll 4\sigma_\text{SB}\varepsilon\,\tilde{T}^3$) the temperature $T_\gas^\-$ does not contribute to the heat transfer problem.
Subsequently, in the $\tau_\rad$ system, the solutions for the particle temperature $T$, given by \eqref{eq:general_solution_temp} and sec. \ref{sec:solution} only depend on the parameters
\begin{align}
	T=T(\varphi_\rad, \vartheta_\rad) \nonumber
\end{align}
(Eqs. \ref{eq:phi} and \ref{eq:theta}).
The photophoretic force \eqref{eq:integral_F} (for directed illumination collapsed to \eqref{eq:f_fm}, i.e. \eqref{eq:f_fm2}) depends on 
\begin{align}
	\mathbf{F}_\text{phot} &= \mathbf{F}_\text{phot}(\varphi_\rad, \vartheta_\rad, \alpha, T_\gas^\-, p) \; . \nonumber
\end{align}
For a contributing $h$, the structure is more complex.

For $\alpha=1$ and $h\ll 4\sigma_\text{SB}\varepsilon\,\tilde{T}^3$ $\rho$, given by \eqref{eq:ratio}, only depends on $\varphi_\rad$ and $\vartheta_\rad$.
For $0<\alpha<1$ $\rho$ will be independent of $p$
\begin{align}
	\rho &= \rho(\varphi_\rad, \vartheta_\rad, \alpha, T_\gas^\-) \; , \nonumber
\end{align}
as shown in \fig{fig:error}, where for four different pairs of $T_\gas^\-$ and $\alpha$ the ratio $\rho$ is plotted over $\varphi_\rad$ and $\vartheta_\rad$.
The dependency of $T_\gas^\-$ and $\alpha$ is only visible on the right half of the plot; for $\varphi_\rad<1$, $\rho$ is basically constant.
Within the parameter range we investigated the deviance for $\varphi_\rad<1$ for different $T_\gas^\-$ and $\alpha$ is only about $10^{-6}$ and therefore negligible.
To deliver a means to classify the accuracy of \eqref{eq:f_fm2} for certain parameters that are within our parameter range, we replotted $\rho$ in \fig{fig:error2} for $\alpha=1$.
The value $\alpha=1$ was chosen because the green area then takes its minimum.
Areas, where $0.98\le\rho\le1.02$ are colored green.
Based on \fig{fig:error2}, a good criterion for a relative error of less than 2\% is
\begin{align}
	\varphi_\rad &\equiv \dfrac{\varepsilon\,I_0\,r_0}{k\,T_\rad} < 1 \; .
\end{align}

\section{CONCLUSION}

Using previous formulae (e.g. by \cite{Hidy1967}) to calculate the longitudinal photophoretic force on a particle with its temperature largely differing from the gas temperature (e.g., due to high $I_0$)  can lead to large errors of several orders of magnitude. As shown in this paper, a new approximation has to be considered for this case in the free molecular flow regime ($Kn \gg 1$). 
If the heat transfer coefficient $h$ (given by \eqref{eq:h}) can be neglected ($h\ll 4\sigma_\text{SB}\varepsilon\,\tilde{T}^3$), the best description of the photophoretic force which still allows analytical treatment in applications is (\eqref{eq:f_fm2})
\begin{align*}
	\mathbf{F}_\text{phot} &\simeq-\frac{\pi}{3} \, \alpha \, \alpha_\text{m} \frac{p}{\sqrt{\overline{T_\gas^\+}\,T_\gas^\-}} \, r_0^2 \, 	\frac{I\,J_1}{\frac{k}{r_0}+4\sigma_\text{SB}\varepsilon\,T_\bb^3} \, \mathbf{e}_z
\end{align*}
with
\begin{align*}
	\overline{T_\gas^\+} &= T_\gas^\-+\alpha\left(T_\bb-T_\gas^\-\right) \\
	T_\bb &= \sqrt[4]{\frac{I_0}{4\sigma_{\text{SB}}}+T_\rad^4} \; .
\end{align*}
The relative error of this equation is very low compared to previous approximations. The average relative error for particles up to a cm is 1\%, for particle sizes up to 1 m it is 3\%.
We provided an error map to assess the relative error depending on the chosen parameters.
For heat transfer coefficients that are comparable or larger than $4\sigma_\text{SB}\varepsilon\,\tilde{T}^3$, the calculation of the photophoretic force follows \eqref{eq:f_fm} using the relations \eqref{eq:kappa_and_psi}.


\section{ACKNOWLEDGMENTS}
This work was funded by DFG 1385.

\appendix
\renewcommand{\theequation}{A.\arabic{equation}}
\renewcommand{\thefigure}{A.\arabic{figure}}
\renewcommand{\thesection}{\Alph{section}}
\renewcommand{\thesubsection}{\thesection.\arabic{subsection}}


\section{SUPPLEMENTARIES}
\subsection{Average}
An average of a physical variable $X$ connected to the gas is given by the integral
\begin{align}
	\overline{X^{\-/\+}} &= \int\limits_{\-/\+}\mathrm{d}^3v\, \sigma^{\-/\+} \, X^{\-/\+} \; . \label{eq:average} 
\end{align}

\subsection{Linearization of $T_\gas^\+$}
Linearization of the square root in \eqref{eq:FphotBaseApprox1} to its first order at $\overline{T_\gas^\+}$
\begin{align}
	\sqrt{T_\gas^\+} = \sqrt{\overline{T_\gas^\+}} +\frac1{2\overline{T_\gas^\+}}\left( T_\gas^\+ - \overline{T_\gas^\+} \right) + \mathcal{O}\left(\left( T_\gas^\+ - \overline{T_\gas^\+} \right)^2 \right) \label{eq:linearization}
\end{align}

\subsection{Legendre polynomials' orthogonality relation}
\begin{align}
	\int_{-1}^{1} P_\nu(x)\, P_\lambda(x)\dd x = \frac2{1+2\nu}\delta_{\nu\lambda} \label{eq:orthogonality}
\end{align}

\subsection{Known surface temperature}
For a known surface temperature, the function $\sqrt{T_\gas^\+}$ can be expanded into a Legendre series
\begin{align}
	\sqrt{T_\gas^\+(r_0,\zeta)} &= \sum\limits_{\nu=0}^{\infty} C_\nu \, P_\nu(\cos\zeta) \; . \label{eq:newApproximation_Tausurface}
\end{align}
Then, \eqref{eq:integral_F} yields
\begin{subequations}
	\begin{align}
		\mathbf{F}_\text{phot} &= -\frac{2\pi}{3} \, \alpha_\text{m} \, \frac{p}{\sqrt{T_\gas^\-}} \, r_0^2 \, C_1 \, \mathbf{e}_z \; . \label{eq:FphotBaseApprox2}
	\end{align}
	$C_1$ \footnote{$C_\nu = \frac{2\nu+1}{2} \int_{-1}^{1} P_\nu(x)\, \sqrt{T_\gas^\+(r_0,x)}\dd x$} can be approximated for $\alpha=\text{const.}$ along \eqref{eq:thermal_accommodation} by the linear term
	\begin{equation}
	C_1 \approx \frac12 \left( \sqrt{T_\gas^\-+\alpha\left(T_{\max}-T_\gas^\-\right)} - \sqrt{T_\gas^\-+\alpha\left(T_{\min}-T_\gas^\-\right)}\right) \; . \label{eq:newApproximation_coefficient}
	\end{equation}
	\label{eq:newApprox2}
\end{subequations}

\begin{figure}[htbp]
	\includegraphics[trim=5 15 195 0, clip, width=\columnwidth]{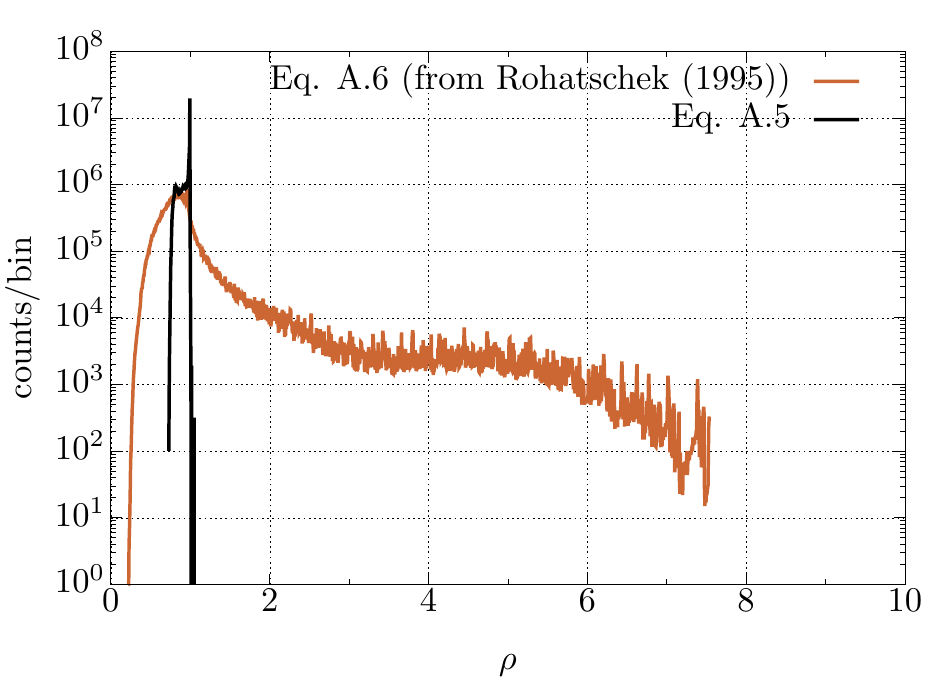}
	\caption{
		Parameter sweep histogram for $63\cdot 10^6$ parameter combinations.
		The parameter sweep intervals are given in \tab{tab:parameter_config}.
		$\rho$ (\eqref{eq:ratio}) is the ratio of the corresponding approximation of the photophoretic force and the total of the effectively exact numerical result given by \eqref{eq:integral_F}, where the surface temperatures were numerically obtained with COMSOL.
		The bin size is 0.005 (0.5\%).
	}
	\label{fig:quality2}
\end{figure}

\cite{Rohatschek1995} and others also use the easier equation
\begin{align}
	\mathbf{F}_\text{phot} &= -\frac{\pi}{6} \, \alpha \, \frac{p}{T_\gas^\-} \, r_0^2 \, \left(T_{\max} - T_{\min}\right) \, \mathbf{e}_z \; . \label{eq:oldApprox2}
\end{align}
\eqref{eq:newApprox2} and \eqref{eq:oldApprox2} will be compared in \fig{fig:quality2} and \tab{tab:statistics2}.
\begin{table*}[!h]
\centering
\caption{
	Statistical properties of the ratio $\rho$ for selected approximations for the photophoretic force arising from directed illumination.
	A parameter sweep of $63\cdot 10^6$ parameter combinations was performed along the parameter intervals given in \tab{tab:parameter_config}.
	 }
\label{tab:statistics2}
\begin{tabular}{$C{4.5cm} >{\bfseries}^C{1cm}>{\bfseries}^C{1cm} ^C{1cm}^C{1cm}>{\itshape}^C{1cm}}
\toprule
approximations for longitudinal \textit{fm}-photophoresis & min & max & mean & median & STD \\
\midrule
\eqref{eq:oldApprox2} (from \cite{Rohatschek1995})			& 0.22 & 7.54 & 0.99 & 0.85 & 0.67 \\
\rowstyle{\bfseries} \eqref{eq:newApprox2}					& 0.73 & 1.05 & 0.93 & 0.96 & 0.07 \\ \bottomrule
\end{tabular}
\end{table*}

\subsection{Units free notation in spherical coordinates}\label{eq:laplace}
The three units free coordinates $(\lambda,\xi,\zeta)$ are in the set $[0,1]\times[0,2\pi]\times[0,\pi]$.
The transformation is given by \eqref{eq:unitsFreeVars}.
The variables and operators in the two coordinate systems $(\lambda,\xi,\zeta)$, denoted by a tilde, and $(r,\xi,\zeta)$ relate as below:
The Laplace operator reads
\begin{equation}
	\Delta = \frac1{r_0^2}\tilde{\Delta} \; ,
\end{equation}
the unit source $q$ is
\begin{equation}
	q = \frac1{r_0}\,\tilde{q} \, ,
\end{equation}
the measure $\mathrm{d}V$ is
\begin{equation}
	\dd V = r_0^3\,\widetilde{\mathrm{d}V} \, .
\end{equation}


\begin{table}[!h]
	\centering
	\caption{\label{tab:table2}Notation.}
	\begin{tabular}{ >{$}r<{ $} L{0.75\columnwidth}  }\toprule		
		\text{variable} & meaning 	\\
		\midrule
		\mathbf{A},\, A=\|\mathbf{A}\| & a vector and its total\\
		\mathbf{r} = (r,\xi,\zeta) & spherical coordinates (\fig{fig:sphere_with_Trot}), $r$ in m \\
		\partial V & border of the volume V (point set), i.e. $r=r_0$ for the sphere \\
		\mathrm{d}\mathbf{A} & surface element vector \\
		\sigma(\mathbf{r},\mathbf{v},t) & gas molecule density with normalization $\int\sigma(\mathbf{r}, \mathbf{v}, t)\dd^3\mathbf{r}\dd^3\mathbf{v} = N$ (gas molecule count) \\
		n(\mathbf{r},t) & spatial gas molecule density, $n(\mathbf{r},t)=\int\sigma(\mathbf{r}, \mathbf{v}, t)\dd^3\mathbf{v}$ \\
		\mathbf{v}(\mathbf{r},t) & gas molecule velocity, in $\mathrm{m~s^{-1}}$ \\
		\overline{v} & mean gas molecule speed $v=\|\mathbf{v}\|$ along \eqref{eq:average}, in $\mathrm{m\,s^{-1}}$\\
		T(r,\xi,\zeta)		& particle temperature, in K	\\
		T_{\max}, T_{\min} & maximum/minimum particle surface temperature \\
		T_\gas^{\+/{\ominus}}	& gas temperature for velocity half-spaces $\mathbf{n}\cdot\mathbf{v}>0$ and $\mathbf{n}\cdot\mathbf{v}<0$	\\
		T_\rad	& temperature of radiation field	\\
		T_\text{bb} & black-body temperature (\eqref{eq:blackBodyTemp}) \\
		m_\text{g}		& mass of a gas molecule, in kg \\
		k_\text{B}		& Boltzmann constant \\
		p				& gas pressure, in Pa \\
		\alpha, \alpha_\text{m} & thermal and momentum accommodation coefficient \\
		J_0 = 0.25 & asymmetry factor \\
		J_1 = 0.5 & asymmetry factor \\
		k				& thermal conductivity of suspended particle, in $\mathrm{W~m^{-1}~K^{-1}}$	\\
		r_0				& radius of spherical particle suspended in gas \\
		\mathbf{n}		& normal vector of a surface \\
		h				& heat transfer coefficient, in $\mathrm{W~m^{-2}~K^{-1}}$ (\eqref{eq:h}) \\
		I				& effective intensity $\varepsilon\, I_0$, in $\mathrm{W~m^{-2}~K^{-1}}$ \\
		\varepsilon		& (mean) emissivity \\
		\sigma_\text{SB}& Stefan-Boltzmann constant \\
		l & mean free path of a gas molecule, in m \\
		\Kn & Knudsen number, $\Kn=l/r_0$ \\
		q				& normalized source function (\eqref{eq:heatEQ}), in $1/\mathrm{m}$ \\
		\mathbf{F}_\text{phot}	& photophoretic force \\
		\rho & ratio of an approximation and the total of \eqref{eq:integral_F}\\
		A_\nu, B_\nu, C_\nu, q_\nu & expansion coefficients ($\nu\ge 0$) \\
		
		\bottomrule
	\end{tabular}
\end{table}

\section*{References}






\bibliographystyle{model5-names}\biboptions{authoryear}


\bibliography{references}

\end{document}